\documentclass[twocolumn,prb,showpacs,aps,floats,floatfix]{revtex4}
\usepackage{graphicx}
\begin{document}
\title{Destruction of diagonal and off-diagonal long range order by
disorder in two-dimensional hard core boson systems}
\author{K. Bernardet, G. G. Batrouni}
\affiliation{
Institut Non-Lin\'eaire de Nice, Universit\'e de Nice--Sophia
Antipolis, 1361 route des Lucioles, 06560 Valbonne, France}
\author{M. Troyer}
\affiliation{Theoretische Physik, Eidgen\"ossische Technische Hochschule
Z\"urich, CH-8093 Z\"urich, Switzerland }
\author{A. Dorneich}
\affiliation{Institut f\"ur Theoretische Physik, Universit\"at
W\"urzburg,
97074 W\"urzburg, Germany}
\date{\today}


\begin{abstract}
We use quantum Monte Carlo simulations to study the effect of
disorder, in the form of a disordered chemical potential, on the phase
diagram of the hard core bosonic Hubbard model in two dimensions.  We
find numerical evidence that in two dimensions, no matter how weak the
disorder, it will always destroy the long range density wave order
(checkerboard solid) present at half filling and strong nearest
neighbor repulsion and replace it with a bose glass phase. We study
the properties of this glassy phase including the superfluid density,
energy gaps and the full Green's function. We also study the
possibility of other localized phases at weak nearest neighbor
repulsion, i.e. Anderson localization. We find that such a phase does
not truly exist: The disorder must exceed a threshold before the
bosons (at weak nn repulsion) are localized. The phase diagram for
hard core bosons with disorder cannot be obtained easily from the soft
core phase diagram discussed in the literature.
\end{abstract}
\maketitle

\section{Introduction}

The two dimensional bosonic Hubbard model has been the subject of
intense interest these past years because it is thought to capture
many of the important qualitative features of two dimensional
superconductors and superfluids at very low temperature. For example,
Helium atoms adsorbed on a surface\cite{ggb1} can clearly be described
by bosons moving in a two dimensional environment. It is then natural
to examine the role of disorder in localizing the bosons and producing
exotic phases such as a bose glass or a normal fluid at zero
temperature.  In the case of soft core bosons with contact repulsion,
the bose glass phase was predicted and studied
theoretically\cite{mfisher} and subsequently verified
numerically\cite{ggb3,nandini}.

Another reason for the increased interest in disordered bosonic
systems is a set of fascinating experiments on the
superconducting-insulating transition suggesting the possibility of a
universal conductance right at the
transition\cite{orr,haviland,jaeger,lee,hebard,geerligs,dynes}.
Several ideas, based on disordered bosonic Hubbard models, have been
suggested\cite{mfisher} to explain these results. Extensive numerical
simulations\cite{ggb1,wallin1,wallin2} appear to support these ideas
qualitatively, although the numerical values of the conductance are
not in agreement.

The question of existence of a normal conducting state at zero
temperature has regained momentum with recent experimental
discoveries\cite{metal}. Attempts to explain this phase proceed via
models of disordered bosons, see for example \cite{das,igor} and
references therein.

Yet another reason to study bosons in external potentials (random or
otherwise) are the recent fascinating experiments on atomic
Bose-Einstein condensates on optical lattices\cite{opticlatt}. In many
cases, such as this one, the relevant bosonic Hubbard model is the
soft core one in others it is the hard core that is of interest.  It
is therefore interesting and important to expose and understand some
of the important differences between these two cases.

The paper is organized as follows. In Sec. \ref{sec:bh} we will first
present the hard core boson Hubbard model, our simulation algorithm and
measurements.  Then, in Sec. \ref{sec:pd} we will first review the
phase diagram of the clean model before presenting our results on the
disordered model at strong and weak near neighbor
repulsions. Conclusions and comments are in section {\bf IV}.

\section{The Boson Hubbard model}
\label{sec:bh}
The hard core boson Hubbard Hamiltonian is given by
\begin{eqnarray}
H=&-&t\sum_{\langle {\bf i},{\bf j}\rangle}(a_{{\bf
i}}^{\dagger}a_{{\bf j}}+ a_{{\bf j}}^{\dagger}a_{{\bf i}}) - \sum_{i}
\mu_{\bf i} n_{{\bf i}}
\nonumber \\
&+&V_{1}\sum_{\langle {\bf i},{\bf j} \rangle}
    n_{{\bf i}} n_{{\bf j}}
\label{hub-ham}
\end{eqnarray}
$a_{\bf i}$ ($a_{\bf i}^{\dagger}$) are destruction (creation)
operators of hard--core bosons on site ${\bf i}$ of a two dimensional
square lattice, and $n_{\bf i}$ is the boson number at site ${\bf i}$
while $\mu_{\bf i}$ is the site dependent chemical potential. This is,
therefore, a site dependent energy which models the disorder in the
system. In the absence of disorder, $\mu_{\bf i}$ becomes the normal
chemical potential, $\mu$. There are other ways to model disorder. For
example, one can have bond dependent hopping parameter ($t_{\bf i,j}$),
or near-neighbor interaction ($V_{1,{\bf i,j}}$). It is thought, though
not fully demonstrated, that these possibilities fall in the same
universality class.  The hopping parameter is chosen to be $t=1$ to
fix the energy scale.  $V$ is the near neighbor
interaction.

To characterize the different phases, we need to measure several
physical quantities. A superfluid phase is characterized by the
absence of long range density order and a non-vanishing superfluid (SF)
density. The SF density, $\rho_s$ is given by
\begin{equation}
\rho_s = \langle W^2 \rangle /2t\beta,
\label{rhos}
\end{equation}
where $W$ is the winding number of the phase of the boson
wave function in one of the two spatial dimensions\cite{ggb3,ggb5}
and $\beta=1/kT$. Long range density
order (such as in the checkerboard solid) is characterized by the
density-density correlation function, $c({\bf l})$, and the structure
factor, $S({\bf q})$, its Fourier transform. They are given by
\begin{eqnarray}
c({\bf l})&=&\langle n_{{\bf j}+{\bf l}} n_{\bf j} \rangle
\nonumber \\
S({\bf q}) &=&  \sum_{\bf l} e^{i {\bf q} \cdot {\bf l}} c({\bf l}),
\label{corrfct}
\end{eqnarray}
where $n_{{\bf j}}$ is the occupancy at site ${\bf j}$. In the presence
of long range order, $S({\bf q})$ will diverge with the system size
for a given ordering momentum, ${\bf q^\star}$, which characterizes
the ordered phase. For example, for checkerboard order, ${\bf
q^\star}= (\pi,\pi)$.

Two other very useful quantities are the equal time Green's function
\begin{equation}
G(|{\bf j}-{\bf i}|)=\langle a_{{\bf j}} a^{\dagger}_{{\bf i}} \rangle,
\label{eqtime}
\end{equation}
and the Green's function in imaginary time,
\begin{equation}
G(\tau)=\langle a_{{\bf i},\tau} a^{\dagger}_{{\bf i},0} \rangle.
\label{time}
\end{equation}
In the superfluid phase, $G(|{\bf j}-{\bf i}|)$ saturates at a nonzero
value for large separations, while $G(\tau)$ tends to zero
exponentially thus yielding the quasiparticle excitation energy
spectrum. In the simulations, $G(|{\bf j}-{\bf i}|)$ was measured
along the lattice axes.

In the presence of disorder, we need to average over realizations of
disorder in addition to the usual statistical average for a given
realization. The number of realizations we used depended on the size
of the system but is typically a few hundred.  We do our simulations
using the stochastic series expansion (SSE) algorithm with worm
updates\cite{SSE}. This
algorithm is numerically exact without any discretization error. In
addition it uses non-local updates, hence even large systems can be
sampled efficiently.

\section{The Phase Diagram}
\label{sec:pd}

The phase diagram of the bosonic Hubbard model with finite contact
repulsion (i.e. soft core) and no nearest neighbor (nn) repulsion has
been studied
extensively in one and two dimensions both with and without
disorder. In the absence of disorder, for incommensurate particle
fillings, one always has a superfluid phase. This phase disappears at
commensurate fillings ($\rho=1,2,3...$) when the onsite repulsion is
large enough\cite{mfisher}. The resulting phase is an incompressible
Mott insulating phase which takes the form of
lobes\cite{mfisher,nandini,ggb7} in the ($t/V_0,\mu/V_0$) plane, where
$V_0$ is the onsite repulsion. This phase is gapped, there is a
substantial energy cost (increase in the chemical potential) for
adding a particle onto a commensurate phase. It was
argued\cite{mfisher} that in the presence of any amount of disorder, a
new compressible insulating (i.e. localized) phase, the bose glass, is
produced at incommensurate fillings and that for strong enough
disorder the gapped phase disappears entirely. This was subsequently
confirmed numerically\cite{ggb3,nandini,ggb2,kanwal,heiko}.

\begin{figure}
\includegraphics[width=7cm]{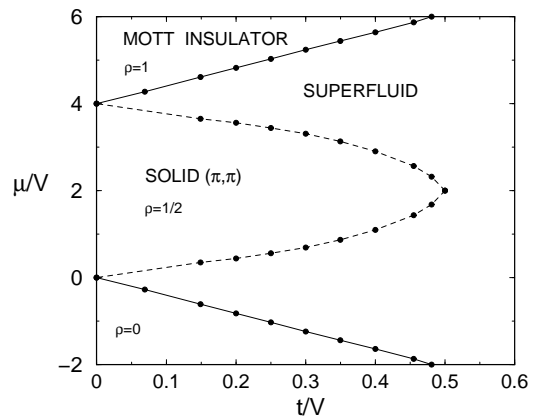}
\caption{The phase diagram of the hard core bosonic Hubbard
model in the absence of disorder ($8\times8$, $\beta=14$). Dashed
lines indicate first order transitions, continuous lines second order
transitions.}
\label{diag_phase_clean}
\end{figure}

The picture changes for hard core bosons with near neighbor repulsion,
$V$. The bosons still form a superfluid for incommensurate particle
filling and $V$ not too strong. Also, at full filling, the bosons
are always frozen into a Mott insulator since hopping to a neighbor
would produce double occupancy which is strictly forbidden. At half
filling, increasing $V$ eventually freezes the bosons into an
incompressible gapped checkerboard solid: Alternate sites are occupied
since the presence of a neighbor costs too much energy and there is a
big energy cost (gap) to add a particle. The phase diagram is shown in
Fig.~\ref{diag_phase_clean}.

\subsection{Strong Near Neighbor Repulsion}

We now introduce disorder in the form of a random site dependent
chemical potential, $\mu_{\bf i}=\mu + \delta_{\bf i}$ where the
disorder, $\delta_{\bf i}$ is uniformly distributed between $\pm
\Delta$.  $\Delta$ is a tunable parameter characterizing the strength of
disorder.  The first question we want to address is the strength of
the disorder necessary to destroy the checkerboard solid phase and
what new phase is produced.  One can try to answer this question with
a simple argument based on energy balance (Imry-Ma). We start at half
filling with a perfect checkerboard solid and introduce the site
disorder. Suppose that at an empty site there is, due to $\mu_{{\bf
i}}$, a deep potential well which pulls in a neighboring boson. This
boson will now have near neighbors which it will try push away to
rearrange its neighborhood in a local checkerboard solid which will,
consequently, have a mismatch at its boundary with the original
checkerboard.  The likelihood of this happening depends on the
disorder and dimensionality. The energy cost, in $d$ dimensions, due
to the mismatch at the boundary scales like $L^{d-1}$ for a region of
length $L$. On the other hand, the energy gained by the bosons by
falling into locally favorable energy wells scales like $L^{d/2}$. For
$d=1$, disorder is relevant, no matter how weak it is, it always
destroys the solid order. For $d=3$ or more, the energy cost outweighs
the gain and the system maintains checkerboard order. The $d=2$ case
is marginal since both, cost and gain, scale like $L$. Typically, in
such marginal cases, the conclusion is that disorder will indeed
destroy long range order but just barely. The correlation length,
$\xi$, is very long and the system size should be even larger to see
the effect.

Numerically, for strong disorder ($\Delta/V=2$) we can easily see that
indeed the gapped checkerboard solid is destroyed on lattices as small
as $L=12$ and is replaced by a compressible, glassy insulator with no
energy gap.

Since for very weak disorder, the system size needed to see the
destruction of solid order is too large for us to simulate, we resort
to finite size scaling for the intermediate disorder case. Although
this does not demonstrate directly the validity of the Imry-Ma
argument (for which very weak disorder is needed) we believe that the
results we will present are qualitatively similar to what happens in
the very weak disorder case.

\begin{figure}
\includegraphics[width=7cm]{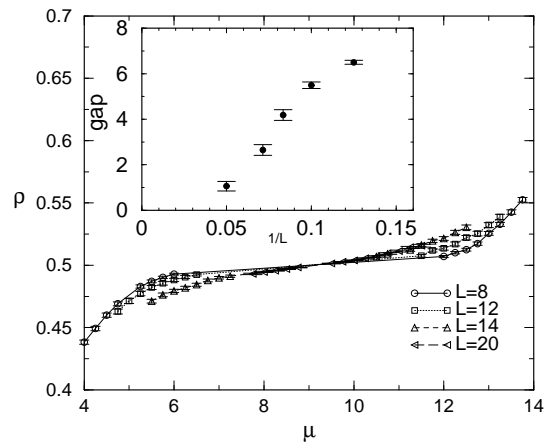}
\caption{$\rho$ versus $\mu$ for $\Delta/V=1$ and different systems sizes
showing the shrinking gap. Inset: The average gap versus
$L^{-1}$. $\beta=14$ for $L=8, 12, 14$ and $\beta=20$ for $L=20$}
\label{rho_mu}
\end{figure}

\begin{figure}
\includegraphics[width=7cm]{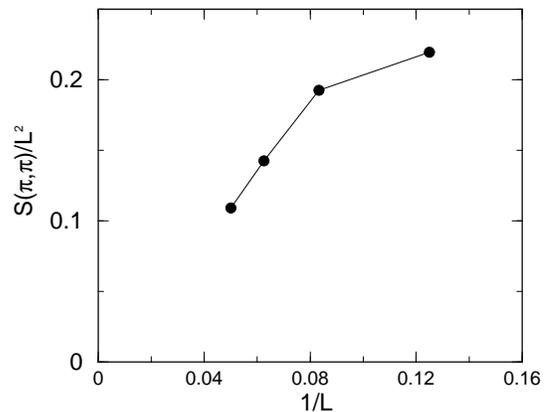}
\caption{$S(\pi,\pi)$ versus $1/L$ for $\Delta/V=1$. $\beta=14$ for
$L=8,10,12,14$ and $\beta=20$ for $L=20$.
Long range order seems to disappear for $L \geq
30$.}
\label{Spipi}
\end{figure}

In Figure~\ref{rho_mu} we show the density, $\rho$, as a function of
the chemical potential, $\mu=\langle \mu_{\bf i}\rangle$, for $L=8,
12, 14$ ($\beta=14$) and $L= 20$ ($\beta=20$) and $\Delta/V=1$.  For
$L=12, 14$ and $20$, we average $100$ disorder realizations, for $L=8$
we did $400$. We see that the incompressible gapped region
($\kappa=\partial \rho / \partial \mu=0$) gets smaller as $L$
increases but does not quite reach zero. The inset shows the average
gap size versus $L^{-1}$. The average gap size was obtained by
calculating the $\rho,\mu$ curve for {\it each} realization, which
yields the gap size per realization which we then average. An
alternative method (which washes out important statistical
information) is to calculate the average $\rho,\mu$ curve and use it
to calculate the average gap.  We see clearly that the gap tends to
zero for a finite, but large, $L$. This suggests that, for these
values of $V$ and $\Delta$, the gap will disappear by $L\approx
30$. Since $L$ should be greater than $\xi$ to observe the destruction
of the checkerboard order, we estimate from this that $\xi \sim
30$. In Figure~\ref{Spipi} we show $S(\pi,\pi)$ as a function of
$L^{-1}$. This, again, shows that the checkerboard order is destroyed
for systems larger than about $L=30$.

\begin{figure}
\includegraphics[width=7cm]{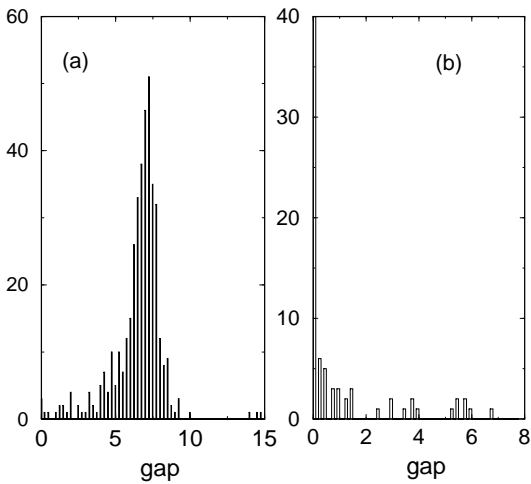}
\caption{Gap distribution for different realizations for
$L=8$ ($\beta=14$, $400$ realizations) (a) and $L=20$ ($\beta=20$,
$100$ realizations) (b). $V=\Delta=4.5$. As lattice size increases,
the distribution gets very wide and peaked at $0$.}
\label{histo_L8L20}
\end{figure}

To elaborate this further, we show in Fig.~\ref{histo_L8L20} the
distributions of the gap sizes for different disorder realizations for
$L=8,20$. We see that for $L=8$, the distribution is quite narrow and
peaked at a nonzero value. However, for $L=20$, the distribution is
very wide and in fact peaked at zero indicating that the most probable
value for the gap is zero. In this case, it is incomplete to discuss
the ``average'' of the gap, which is still non-zero.

\begin{figure}
\includegraphics[width=7cm]{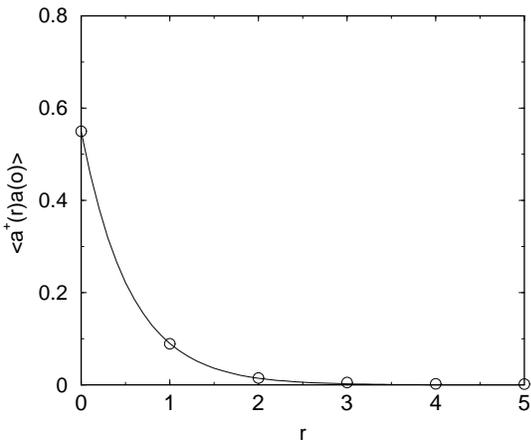}
\caption{Equal time Green's function as a function of distance for
$L=10, \beta=20, V=4.5$, and $\Delta/V=1$. The solid line is a fit of
the form: $y=A_0 ({\rm exp}(-(L/2-x)/A_1)+{\rm exp}((L/2-x)/A_1))$
with $A_0=6.4\times 10^{-5}$ and $A_1=0.552$.  }
\label{equaltime-g}
\end{figure}

In order to characterize further the compressible phase which replaces
the checkerboard solid, we study the behavior of the Green's function,
both for equal and unequal imaginary times. For example, in
Figure~\ref{equaltime-g} we show the equal time Green's function for
$V=4.5$, $L=10$ and $\rho \sim 0.56 $. We see that the Green's
function goes to zero and is very well fit by an exponential (in fact
a hyperbolic cosine to account for the periodic boundary
conditions). This is further evidence that there is no superfluid in
this phase, it is an insulator.

\begin{figure}
\includegraphics[width=7cm]{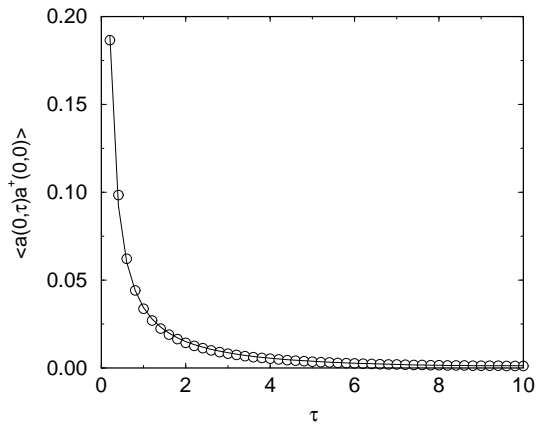}
\caption{The Green's function, $G(\tau)$ as a function of imaginary time
separation, same parameters as Fig. 5. The solid line is a fit of the
form: $G(\tau) =A_0(\tau^{A_1}+(\beta/2 - \tau)^{A_1})+A_2$ with
$A_0=0.04$, $A_1=-1.0$ and $A_2=-0.0065$. With a two parameter fit
(excluding $A_3$) we also get a good fit with $A_0=0.03$
and$A_1=-1.16$.}
\label{time-g}
\end{figure}

The glassy nature of this phase can be seen in the time-dependent
Green's function, Eq.~\ref{time}. In the Bose glass phase, this
quantity is predicted\cite{mfisher} to decay as $G(\tau) \sim 1/\tau$,
which has been verified numerically in a very different
context\cite{heiko}. Figure~\ref{time-g} shows that is also true in
this case. Therefore, the new phase replacing the gapped checkerboard
solid is an ungapped insulating Bose glass phase.

Clearly, it is very difficult to examine these issues with smaller
couplings and disorder: The correlation length will be even longer and
much larger sizes would be needed. However, for moderate disorder, the
above results demonstrate that, whereas it might appear on a finite
lattice that the gapped solid phase is still present, finite size
scaling clearly shows the gap to disappear on large enough systems. We
may conclude from this that the Imry-Ma argument holds and that
disorder, no matter how weak, will produce a glassy, compressible,
ungapped insulating phase at strong near neighbor couplings.

\subsection{Weak Near Neighbor Repulsion}

We now consider the question of what happens when the near neighbor
repulsion is decreased and only the hard core and disorder
interactions remain. For soft core bosons with no nn interaction, a
re-entrant behavior was observed for $\rho_s$ as a function of the
contact repulsion both in one\cite{ggb3} and two\cite{nandini}
dimensions. In other words, for fixed disorder strength, as the onsite
repulsion is increased from zero, the superfluid density is at first
zero, then at some intermediate value of $V_0/t$ the bosons delocalize
and $\rho_s$ takes on a finite value, then for large enough $V_0/t$
({\it i.e.}  approaching the hard core limit) the bosons are localized
again. In one dimension this happens for any amount of disorder, but
in two dimensions was only reported\cite{nandini} at
$\Delta/t=6$. This was taken as confirmation of the phase diagram
presented in reference~\onlinecite{mfisher} where the bosons were
argued to be always localized, even by weak disorder, when $V_0/t$ is
very large.

The phase diagram in the $(t/V_0, \mu/V_0)$ plane\cite{mfisher} should
however be interpreted with care especially if we want to consider the
hard core limit. It is a phase diagram at constant {\it finite}
disorder $\Delta/V_0$ and thus $t/V_0\to 0$ means that $\Delta/t\to
\infty$, and any arbitrarily weak disorder in units of $V_0$ becomes
infinitely strong in the hard core limit.

\begin{figure}
\includegraphics[width=7cm]{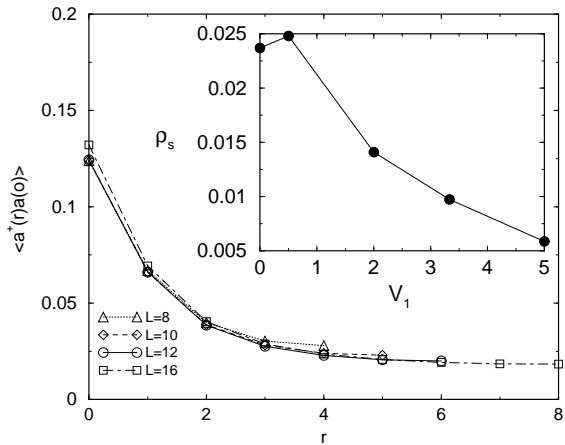}
\caption{The Green's function versus $r$ for $V=0, \Delta=5, \beta=24,
\rho=0.1$ and $L=8,10,12,16$ showing that the system is superfluid.
The inset shows $\rho_s/\rho$ versus $V$ for a $12X12$
system. $\Delta/t=5, \beta=24, \rho=0.1$ and $250$
realizations. $\rho_s/\rho$ vanishes only for large $V$. The
particle density is $\rho\approx 0.5$.}
\label{rhos_V1}
\end{figure}

In fact, figure~\ref{rhos_V1} shows that hard core bosons behave
differently, when a finite disorder $\Delta/t$ is considered.  While
the bosons are localized by weak disorder for large nn coupling, $V_1$
(and densities that are not very small), as this coupling is reduced,
the bosons become superfluid and stay that way even at $V_1=0$. We
verified this at several particle densities. In other words, the naive
expectation based on the soft core phase diagram as presented in
reference~\onlinecite{mfisher} at very large onsite repulsion is not
fulfilled. The soft core phase diagram must be interpreted
appropriately.

\begin{figure}
\includegraphics[width=7cm]{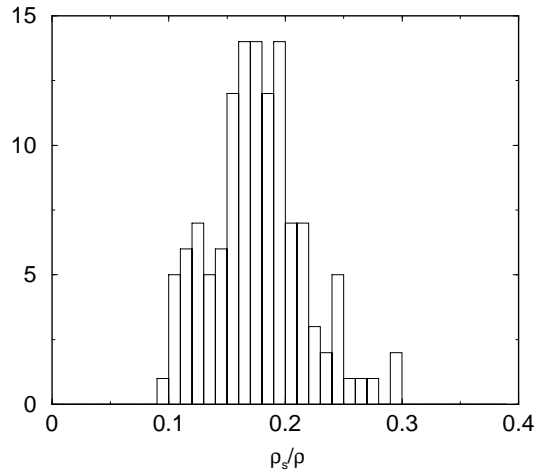}
\caption{Histogram of $\rho_s/\rho$ for the $250$ realizations for
$L=16, V=0, \Delta/t=5, \beta=24$ and $\rho\approx 0.1$.}
\label{histo-rhos}
\end{figure}

As we saw for the gap, in the presence of disorder it is not enough to
consider only average quantities, it is also instructive to consider
their distribution. In Figure~\ref{histo-rhos} we show the histogram
of $\rho_s/\rho$ for the $250$ realizations for $L=16, V=0, \Delta=5t$
and $\rho\approx 0.1$. We see that for this size the distribution
is well centered around $\rho_s/\rho\approx 0.18$, with the same
behavior observed for smaller sizes: The distribution does not widen
nor does the peak tend to zero. Therefore, even for this large
disorder, the bosons are still not localized for $V=0$.

\begin{figure}
\includegraphics[width=7cm]{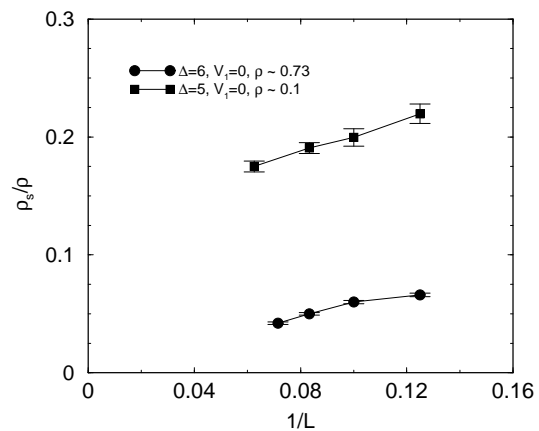}
\caption{Finite size scaling of the superfluid density fraction
$\rho_s/\rho$ for the pure hard-core case $V=0$ for two different
disorder strengths $\Delta=5t$ and $\Delta=6t$, $\beta=20$.}
\label{FSS-rhos}
\end{figure}

\begin{figure}
\includegraphics[width=7cm]{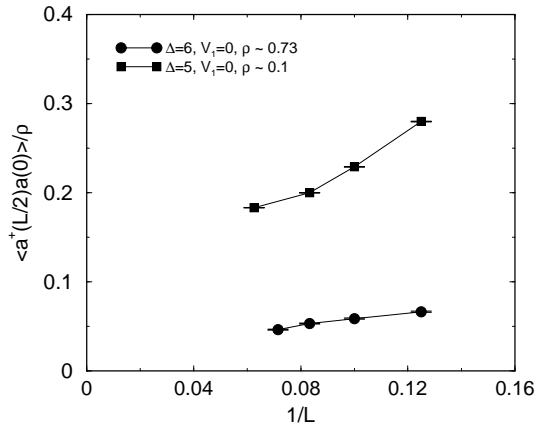}
\caption{Finite size scaling of the equal time Green's function at the
largest distance for two different disorder strengths $\Delta=5t$ and
$\Delta=6t$, $\beta=20$}
\label{FSS-Green's}
\end{figure}

To show that the bosons can be localized at $V=0$ if the disorder
exceeds a threshold value, we show in Fig.~\ref{FSS-rhos} finite size
scaling for $\rho_s/\rho$ for two cases of disorder, $\Delta=5t$, and
$6t$. The case of $\Delta = 5t$ was done with $\rho=0.1$ and clearly
shows the system to be superfluid as as $L\to \infty$. This is also
shown in the equal time Green's function, Fig.~\ref{FSS-Green's}. On
the other hand for $\Delta=6t$ we did the simulation at
$\rho\approx0.73$ and we find that both $\rho_s/\rho$ and the Green's
function vanish as $L\to \infty$. We also verified this for
$\rho\approx 0.5$ and $0.1$. We conclude that the disorder has to
exceed a threshold value ($\approx 6t$) in order to localize the
bosons for {\it weak} nn repulsion. Clearly, in the no-hopping limit,
$t\to 0$, the critical value of the disorder vanishes in agreement
with reference
\onlinecite{mfisher}.

When the disorder is strong enough to localize the bosons at weak
coupling and consequently also at strong coupling (since even very
weak disorder can accomplish that), the question arises as to whether
there is re-entrant behavior, as observed for soft core
bosons\cite{ggb3,nandini}. In other words, will the bosons go from
being localized at weak coupling to being delocalized at intermediate
values and then relocalize at strong coupling? A finite size scaling
analysis shows that this is not the case: For the hard core system,
when the disorder is large enough to localize the bosons at weak
coupling, they will stay localized at all couplings.

This means that the ($\mu/V,t/V$) phase diagram is as follows. For
weak disorder the checkerboard solid is completely destroyed and
replaced by a compressible insulating Bose glass phase. This is easy
to understand by applying to the two-sublattice structure of the
checkerboard solid a simple Imry-Ma argument. However, away from half
filling, there is no such structure and the Imry-Ma argument no longer
holds. So, with weak disorder, there will be a superfluid even for
large nn repulsion, when the density is far from $0.5$. In addition,
there is a superfluid for all fillings (except full filling) when the
nn repulsion is small.  For strong disorder, the superfluid phase
everywhere is destroyed and replaced by the compressible Bose glass
phase.

\section{Conclusions and Discussion}

We have used the SSE algorithm to simulate the disordered hard core
bosonic Hubbard model in two dimensions to try to understand the
interplay and competition between interaction and disorder.

Using finite size scaling, we found that the checkerboard solid
present at strong nn interactions for half filling is completely
destroyed by any amount of disorder. This agrees with the simple
Imry-Ma energy balance argument.

Surprisingly, finite size scaling showed that at very weak, even
vanishing, nn coupling, weak (even intermediate) disorder does not
localize the hard core bosons. The disorder strength, $\Delta/t$, must
be at least of the order of $6t$ before the bosons are localized.  We
found this somewhat surprising because soft core bosons were argued to
be always localized\cite{mfisher}, even by weak disorder, in the limit
$t/V_0 \to \infty$ which is often considered to be equivalent to the
hard core case. The soft core phase diagram\cite{mfisher} should be
interpreted carefully, as discussed in section {\bf IIIB}.

Numerically, the bosons were shown to be localized but only one value
of the disorder was presented\cite{nandini}, $\Delta/t=6$, which is
large. Our numerical results here agree with
reference~\onlinecite{nandini}, but show no localization for weaker
disorder.

Another interesting analogy to make is with fermions. In one
dimension, weak disorder localizes both fermions and hard core bosons,
which is not surprising in view of the fact that they are equivalent
in this case. In two dimensions, however, this equivalency is lost,
and the response to disorder is different. Fermions are (marginally)
localized by weak disorder while hard core bosons are not.

Since the hard core bosonic Hubbard model is equivalent to the
spin$-{1 \over 2}$ quantum Heisenberg model, the above results also
hold for this model in the presence of a random external magnetic
field.

\acknowledgments

We acknowledge very helpful conversations with R. T. Scalettar,
H. Rieger. G.G.B. acknowledges support from the Franco-German PROCOPE
program. M.T. was supported by the Swiss National Science
Foundation. The calculations were performed on the Asgard Beowulf
cluster at ETH Z\"urich and the Cray T3E at HLRS (Stuttgart) using a
parallelizing C++ library for Monte Carlo simulations\cite{alea}.

\end{document}